\documentclass[amsmath,amssymb,aps]{revtex4-2}   

\usepackage{graphicx}
\usepackage{hyperref}
\usepackage{xcolor}

\newcommand{\aver}[1]{ \! \left\langle {#1} \right \rangle \!}
\newcommand{\vect}[1]{\boldsymbol{#1}}

\begin{document}

%\preprint{APS/123-QED}

\title{Coherent near-wall structures and drag reduction by spanwise forcing}

\author{Emanuele Gallorini}
 \email{emanuele.gallorini@polimi.it}

\author{Maurizio Quadrio}
 \email{maurizio.quadrio@polimi.it}
\affiliation{%
Department of Aerospace Sciences and Technologies, Politecnico di Milano, via La Masa 34,\\
20156 Milano, Italy 
}%

\author{Davide Gatti}
 \email{davide.gatti@kit.edu}
\affiliation{%
Institute of Fluid Mechanics, Karlsruhe Institute of Technology, Kaiserstraße 10,\\
76131 Karlsruhe, Germany
}%

\begin{abstract}
The effect of streamwise-traveling waves of spanwise wall velocity (StTW) on the quasi-streamwise vortices (QSV) populating the near-wall region of turbulent channels is studied via a conditional averaging technique applied to flow snapshots obtained via Direct Numerical Simulation. The analysis by A. Yakeno, Y. Hasegawa, N. Kasagi, ``Modification of quasi-streamwise vortical structure in a drag-reduced turbulent channel flow with spanwise wall oscillation'', Phys. Fluids 26, 085109 (2014), where the special case of spatially uniform wall oscillation (OW) was considered, is extended to the general case of StTW, which yield both reduction and increase of turbulent skin-friction drag. StTW are found to significantly impact the wall-normal distribution of the vortex population. The conditionally averaged velocity field around the vortices shows that the contributions of the QSV to the quadrant Reynolds shear stresses change significantly during the control cycle. On the one hand, as for OW, the suppression of Q2 events (with upwelling of low-speed fluid away from the wall) dominates the drag-reduction process. On the other, the enhancement of Q2 and also Q4 events (with downwelling of high-speed fluid toward the wall) is related to drag increase. Based on the link identified between the phase changes of the Reynolds stresses and the principal directions of the rate-of-strain tensor induced by the StTW, a predictive correlation for drag reduction by StTW is proposed which uses physically significant parameters to overcome the shortcomings of existing models.
\end{abstract}

\maketitle

\section{Introduction}
\label{sec:intro}

During the last decades, important research efforts have been devoted to the development of active, predetermined flow-control strategies for the reduction of skin-friction drag in turbulent flows and to the physical understanding of drag reduction. Predetermined strategies can achieve sizable net energy savings at the cost of acceptable energy expenditure, with the advantage of lower complexity of the control system, which does not need sensors.

The streamwise-travelling waves of spanwise wall velocity (StTW), subject of the present work, were introduced one decade ago by Quadrio and coworkers \cite{quadrio-ricco-viotti-2009} and are a prominent member in the class of predetermined drag reduction strategies. Recent and comprehensive reviews are available in \cite{leschziner-2020,ricco-skote-leschziner-2021}. StTW can relaminarize an otherwise turbulent flow, achieve large net energy savings, and suit both internal and external flows. They have had successful laboratory implementations starting with Auteri et al. \cite{auteri-etal-2010} who ideated the actuation strategy later employed in \cite{marusic-etal-2021}; they remain effective at large Reynolds numbers \cite{gatti-quadrio-2016}, and have been recently demonstrated to reduce both friction and pressure drag over non-planar walls \cite{banchetti-luchini-quadrio-2020} and to provide large reduction of the drag for an airplane in transonic flight \cite{quadrio-etal-2022} by interacting with the shock wave. However, the working mechanism of StTW remains not fully understood, and the theoretical interest stems from the contrast between the relatively simple action performed by the control --- a periodic lateral motion of the walls --- and the resulting significant reduction in skin-friction drag.

The travelling waves result from a spanwise wall velocity distribution of the form 
\begin{equation}
w_w\left(x, t \right) = A \sin \left( \omega t - \kappa_x x \right)
\label{eq:twlaw}
\end{equation}
where $w_w\left(x, t \right)$ is the spanwise velocity component at the wall, which periodically varies according to the phase $\phi = \omega t - \kappa_x x$, function of both the time $t$ and the streamwise coordinate $x$. The control parameters are the maximum wall velocity $A$, the wavenumber $\kappa_x$ and the angular frequency $\omega$. These, in turn, determine the wavelength $\lambda_x = 2\pi / \kappa_x$, the period $T = 2\pi / \omega$ and the phase speed $c = \omega / \kappa_x$ of the wall velocity wave. The law \eqref{eq:twlaw} contains the two limiting cases of the stationary wave when $\omega=0$ and the spanwise oscillating wall (OW) when $\kappa_x=0$.

The control law \eqref{eq:twlaw} results in the formation of a so-called generalised Stokes layer (GSL) \cite{quadrio-ricco-2011}, a periodic and streamwise-varying crossflow confined within the vicinity of the walls. For StTW tuned to yield drag reduction, the GSL closely resembles the laminar GSL, which for the OW reduces to the solution of the laminar Stokes' second problem. The mean momentum transport within this layer is related to the quasi-stramwise vortices and longitudinal low-speed streaks which are essential in the near-wall turbulence regeneration cycle \cite{robinson-1991b}. Altering this cycle through a manipulation of QSV and streaks is generally recognised as the foundation of the drag-reducing capabilities of the GSL \cite{yakeno-hasegawa-kasagi-2014}. However, the details of the drag reduction mechanisms via StTW remain partially unclear, especially in regard to the interaction between the wall forcing and the near-wall turbulent structures. The recent work \cite{umair-tardu-doche-2022}, for example, explores changes to the Reynolds stresses induced by StTW, but does not provide any dynamical link to structural changes in the flow.

Similarly incomplete is the quest for a predictive correlation, based on physical or empirical arguments, capable to estimate the drag reduction as a function of the control parameters. While the ability to predict drag reduction may have in itself  limited practical appeal, a good prediction formula based on physical quantities provides the added value of hinting at the important quantities in the drag reduction process. 
Most of the available studies only discuss the specific case of the OW, and very few propose predictive correlations. Quadrio and Ricco \cite{quadrio-ricco-2004} suggested drag reduction to be linearly proportional to a parameter that contains both the maximum wall velocity and the oscillation period. Drag reduction physics has been considered by Touber \& Leschziner \cite{touber-leschziner-2012}, who linked the alignment of the near-wall streaks to the Stokes strain at $y^+=10$, where $y$ denotes the wall-normal direction and the superscript $+$ indicates nondimensionalisation in viscous units, i.e. with the friction velocity $u_\tau$ and the fluid kinematic viscosity $\nu$.  When the direction of the local strain rate changes rapidly in time compared to the characteristic streak formation time of $50 \nu / u_\tau^2$ (estimated in Ref. \cite{blesbois-etal-2013} via a global optimum perturbation analysis) or to the integral lifetime of streamwise wall shear fluctuations of $60 \nu / u_\tau^2$ (measured in Ref. \cite{quadrio-luchini-2003} via time-space corrrelations), streaks organisation is disrupted, leading to drag reduction. 
Agostini \emph{et al.} \cite{agostini-touber-leschziner-2014} elaborated further on this aspect, suggesting that a large wall-normal derivative of the velocity vector in the lower part of the Stokes layer is important for drag reduction. In their sequel study \cite{agostini-touber-leschziner-2015}, the OW effect on enstrophy budgets is studied to find that the vortex tilting/stretching of wall-normal vorticity by spanwise velocity fluctuations is linked to the control-induced shear stress variations, while the low-drag state is associated with the enstrophy production term related to the wall-normal vorticity component. While important to understand the drag reduction process, these studies have no predictive value, since they correlate drag reduction with turbulent quantities which are typically unknown {\em a priori}. 

The goal of this work is two-fold. We first extend the conditional analysis of Yakeno, Hasegawa and Kasagi \cite{yakeno-hasegawa-kasagi-2014}, hereinafter indicated as YHK14, to the general case of StTW. Moreover, the limitations of existing predictive drag reduction correlations applied to StTW are discussed, and a new predictive correlation is developed, combining empirical and physical arguments, to specifically address those limitations and to arrive at a satisfactory prediction based upon GSL quantities alone. After describing the simulations carried out for this work in \S\ref{sec:methods}, the paper continues in \S\ref{sec:quadrant} with a quadrant analysis of the Reynolds stresses, and in \S\ref{sec:conditional} with a conditional analysis that connects quadrant changes to structural modifications in the flow. This understanding is then leveraged in \S\ref{sec:dragchanges} to arrive at a formula for the prediction of drag changes that, starting from what is available for the OW, is capable to work well for the more general form of forcing. Lastly, \S\ref{sec:conclusions} contains a brief concluding discussion.

\section{Methods}
\label{sec:methods}

\begin{table}  
\center    
\begin{tabular}{llrrrrrr}
\hline\noalign{\smallskip}
Case & $A^+$ & $\omega^+$ & $\kappa^+$ & $N$ & $Re_b$ & $\Delta U_b^+$ & $\Delta C_f$ (\%) \\
\noalign{\smallskip}\hline\noalign{\smallskip}
Ref &  --  & --     & --   &  90  &  3180  &   --   & - \\
OW1 &  7   & 0.0838 & 0   & 368  &  3674  &  2.48  &  25.2 \\
OW2 &  7   & 0.0251 & 0   & 368  &  3414  &  1.17  &  13.2 \\
TW1 &  7   & 0.0238 & 0.01 & 100  &  3982  &  4.02  &  36.3 \\
TW2 &  7   & 0.12   & 0.01 & 100  &  2930  & -1.25  & -17.8 \\
\noalign{\smallskip}\hline
\end{tabular}
\caption{Parameters for the five simulations, all at $Re_\tau=200$. The table reports on the left columns the control parameters $A$, $\omega$ and $\kappa$,  and the number $N$ of flow fields stored for further analysis. The right columns illustrate the obtained drag reduction, in terms of bulk Reynolds number $Re_b$, the increase of the bulk velocity $\Delta U_b^+$ and the percentage change of the skin-friction coefficient $\Delta C_f$.}
\label{tab:datasets}
\end{table}

Five incompressible Direct Numerical Simulations (DNS) of a fully developed turbulent channel flow are performed to assess how quasi-streamwise vortices change under the action of the oscillating wall and of streamwise travelling waves of spanwise-wall velocity. One DNS provides the baseline reference case, whereas for each control technique two configurations are simulated, one with good performance and the other that performs poorly (smaller drag reduction for OW and drag increase for StTW). Table \ref{tab:datasets} lists the five cases and provides additional information, including drag reduction data, which are all in agreement with existing information. The primary Reynolds number is based upon the friction velocity $u_\tau$ and half the gap $h$ between the plane walls, i.e. $Re_\tau=h u_\tau / \nu$; in all cases it is set at $Re_\tau=200$. All the simulations are run under the constant pressure gradient (CPG) condition \cite{quadrio-frohnapfel-hasegawa-2016}, so that a unique value of $u_\tau$ is available and ambiguity in scaling is avoided. Under CPG, drag reduction implies an increase of the bulk velocity $U_b$, and a corresponding decrease of $C_f$. 

Drag reduction is commonly defined as
\begin{equation}
   R = 1 - \frac{C_f}{C_{f,0}} \, ,
\label{eq:R}
\end{equation}
i.e. as the relative change in skin friction coefficient $C_f = 2 / (U_b^+)^2$ with respect to a reference uncontrolled case (indicated by the subscript 0). StTW yield a drag reduction up to $R \approx 0.45$ at $A^+=12$ and $Re_\tau = 200$ \cite{quadrio-ricco-viotti-2009,hurst-yang-chung-2014,gatti-quadrio-2016}. The corresponding value of $Re$ always needs to be specified along with $R$, since $R$ is an inherently Reynolds-dependent measure of drag reduction. As discussed, for instance, by Gatti \& Quadrio \cite[henceforth also [GQ16]{gatti-quadrio-2016}, $R$ necessarily decreases with $Re$. Similarly to what occurs for drag reduction via riblets \cite{garcia-jimenez-2011-a} or drag increase due to roughness, [GQ16] have shown that StTW produce an upward shift $\Delta U^+$ of the logarithmic portion of the mean velocity profile, and that $\Delta U^+$ is the true Reynolds-independent measure of drag reduction.

The employed DNS solver, developed by Luchini and Quadrio \cite{luchini-quadrio-2006}, solves the incompressible Navier--Stokes equation with spectral discretization in the homogeneous directions and compact fourth-order finite differences in the wall-normal direction. The size of the computational domain is $4\pi h \times 2h \times 2\pi h$ along the streamwise $x$, wall-normal $y$ and spanwise $z$ directions; $N_x=N_z=256$ Fourier modes discretize the homogeneous directions, while $N_y=193$ nodes, distributed along $y$ with an hyperbolic tangent, discretize the wall-normal direction. The spatial resolution in wall units is thus $\Delta x^+ = 6.5$ and $\Delta z^+ = 3.3$ once the additional modes used to remove the aliasing error are factored in; $\Delta y^+$ smoothly varies from $\Delta y^+\approx 0.5$ near the wall to $\Delta y^+\approx 3.7$ at the centerline. (Here and in the following, all the quantities will be expressed in wall units and the superscript $^+$ will be omitted, with the exception of spatial coordinates, e.g. $y^+$).

The initial condition for all simulations is an uncontrolled turbulent flow field: the flow thus experiences a transient phase, which is discarded in the later statistical analysis. To enable data saving at predetermined phases of the control law, the value of the time step is chosen as an integer submultiple of the forcing period, and is such that the maximum CFL remains below unit. Unit CFL is well below the stability limit of the temporal integration scheme, which is a combination of the implicit Crank-Nicolson scheme used for the viscous terms, and a third-order Runge-Kutta scheme for the convective terms. 
%Ref: 0.1178; OW1: 0.0938; OW2: 0.0781; TW1: 0.0974; TW2: 0.1102

The two OW cases, referred to as OW1 and OW2, are at $A=7$, but the oscillation period changes with $T=75$ and $T=250$ respectively. The specific forcing amplitude, chosen to properly compare with YHK14, is near the amplitude at which spanwise forcing provides the largest net benefits. Data are extracted at eight equally spaced phases along the period. Drag reduction of OW1 at 25.2\% is twice that of OW2 at 13.2\%. The two StTW cases are at the same $A=7$ and $\kappa=0.01$, but the oscillation frequency is changed at $\omega=0.0238$ ($T=264$) and $\omega=0.12$ ($T=52$). TW1 yields the largest drag reduction at 36.3\%, whereas TW2 is in the drag-increasing region of the parameter space, so that $C_f$ increases by 17.8\%. Statistics for OW are computed as a function of the phase along the cycle by averaging along the homogeneous directions and phases, with $N=368$ flow fields are stored at 8 phases along 46 oscillation cycles. To compare OW and StTW at the same phase, the correct streamwise coordinate needs to be chosen. Each of the $N=100$ flow fields with StTW contains four forcing wavelengths and thus four points at the same phase. Collecting statistics over a thin streamwise slice, centered at the right phase and five points thick, is used to increase the statistical sample.

\section{Reynolds shear stress analysis}
\label{sec:quadrant}

The friction drag changes induced by the control are reported in Table \ref{tab:datasets} in terms of their time-averaged values. They are quantified by the change in bulk velocity $\Delta U_b = U_b - U_{b,Ref}$, where $U_{b,Ref}=15.90$ is the bulk velocity of the reference case. Since the comparison is run at CPG, $\Delta U_b$ coincides with the vertical shift $\Delta U$ of the mean velocity profile \cite{gatti-quadrio-2016}. 

Under CPG, $U_b$ can be expressed \cite{marusic-joseph-mahesh-2007} via an extended FIK identity \cite{fukagata-iwamoto-kasagi-2002} as the sum of a laminar contribution and a (negative) term produced by turbulent shear stresses:
\begin{equation}
U_b = \frac{Re_\tau}{3} - \int_{0} ^{Re_\tau} \Bigl( 1 - \frac{y}{Re_\tau} \Bigr) \left( -\overline{u'v'} \right) dy^+. 
\label{eq:DUb}
\end{equation} 
where the overbar is the temporal mean, and the prime indicates a fluctuating turbulent quantity according to the Reynolds decomposition.

Since the control action is periodic, the velocity field is decomposed following YHK14 as:
\begin{equation}
u(x,y,z,t) = \overline{u}(y) + \tilde{u}(y,\omega t - \kappa_x x) + u''(x,y,z,t),
\end{equation}
where the $u$ component is taken as example. In the expression above, the three terms on the right-hand-side represent respectively the time-averaged, periodic and random components. The phase-averaged velocity is also introduced as:
\begin{equation}
\aver{u}(y,\phi) \equiv \overline{u}(y) + \tilde{u}(y,\phi) 
\end{equation}
which is computed as the mean over the $N$ samples along $x$ and $z$ directions at a given phase $\phi$, i.e.:
\begin{equation}
\aver{u}(y,\phi) = \frac{1}{N} \sum_{n=0}^{N-1} \frac{1}{L_x}\frac{1}{L_z}\int_{0}^{Lx}\int_{0}^{Lz} u(x,y,z,\phi+2\pi n)dxdz .
\end{equation} 

Owing to the continuity constraint, $\tilde{v}(y,\phi)=0$, and $\overline{u'v'}=\overline{u''v''}$. 

\begin{figure}
\centering
\includegraphics[width=0.8\textwidth]{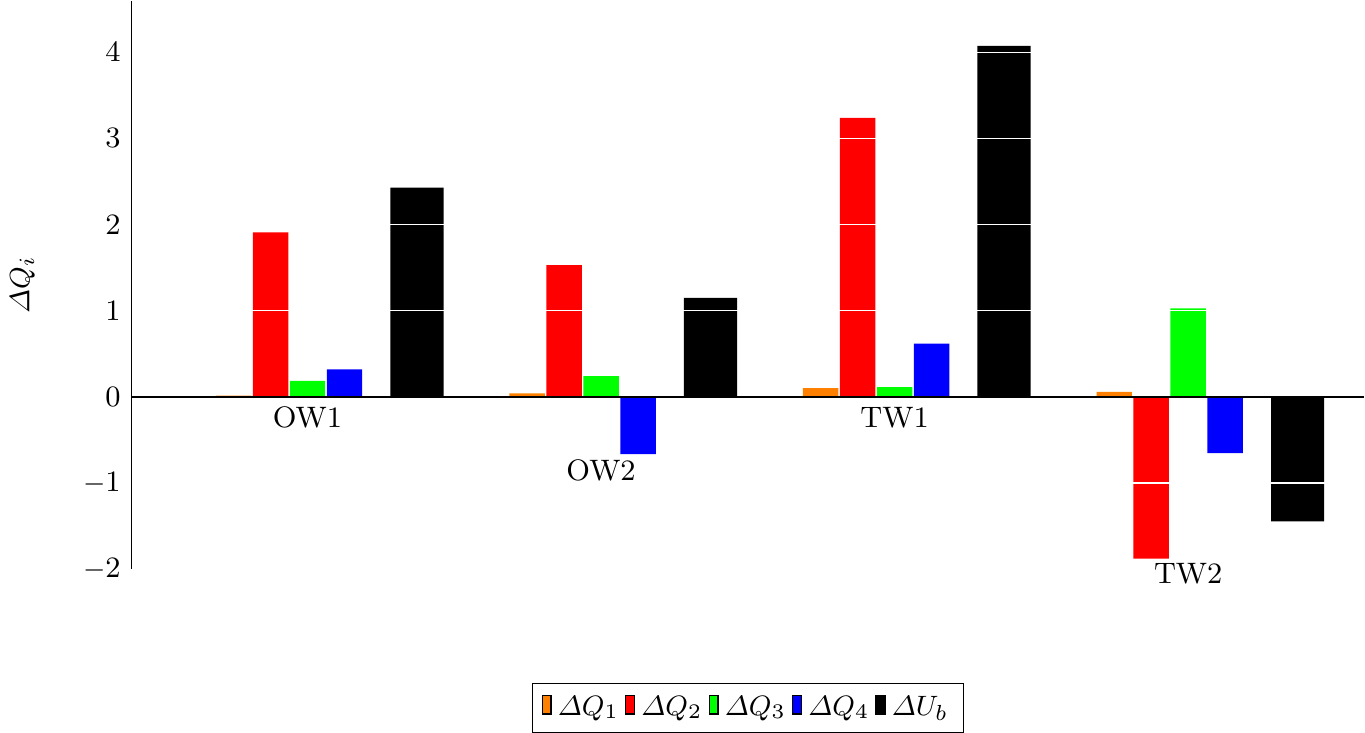}
\caption{Changes $\Delta Q_i$ of the quadrant contributions to $U_b$, from Eq.\eqref{eq:UbFIK}. The sum of the four $\Delta Q_i$ equals the total change $\Delta U_b$ (shown with a black bar) from the uncontrolled case.}
\label{fig:hist}
\end{figure}

The Reynolds shear stresses $\overline{u'v'}$ are commonly assigned to four quadrants, depending on the sign of the fluctuations $u'$ and $v'$: Q1 ($u'>0$ and $v'>0$) and Q3 ($u'<0$ and $v'<0$) events are called interaction modes, Q2 events ($u'<0$ and $v'>0$) are called ejections while Q4 events ($u'>0$ and $v'<0$) are called sweeps \cite{Wallace-etal-1972}. YHK14 applied quadrant analysis to Eq.\ref{eq:DUb}, which is rewritten as:
\begin{equation}
U_b = \frac{Re_{\tau}}{3}+\sum_{i=1}^4 Q_i = \frac{Re_{\tau}}{3} - \sum_{i=1}^4 \int_{0}^{Re_{\tau}} \left( 1-\frac{y^+}{Re_\tau} \right) 
( - \overline{u'' v''} )_i dy^+,
\label{eq:UbFIK}
\end{equation}
where $Q_i$ is the contribution of events from quadrant $i$ to the weighted integral of the Reynolds shear stresses. Since the laminar term remains unchanged between controlled and uncontrolled flows, the change of bulk velocity is the sum of the changes of quadrant contributions to the shear stress: $\Delta U_b = \sum_{i=1}^4 \Delta Q_i$. 

The quadrant analysis was performed by YHK14 for the oscillating wall. Here it is carried out for OW1 and OW2, and then extended to the travelling-wave cases TW1 and TW2. Results are presented in Fig.\ref{fig:hist}. As in the previous study, for the oscillating wall suppression of Q2 events (with the corresponding positive $\Delta Q_2$) is the dominant mechanism which influences drag reduction, while an enhancement of Q4 events shows up in the less effective case OW2 at non-optimal oscillation. Reduction of Q2 events is the dominant mechanism for the travelling waves too: indeed, there is a strong suppression of Q2 events in the drag-reducing TW1 case, whereas in the drag-increasing TW2 case Q2 events are significantly enhanced. TW2 presents other noteworthy features: a negative $\Delta Q_4$ contributes to degrading control performance, similarly to OW2, but this is more than compensated for by the large $\Delta Q_3 > 0$.

\section{Conditional analysis of QSV}
\label{sec:conditional}

The changes $\Delta Q_i$ of the shear stresses can be linked to structural modifications in the flow. In fact, the control affects the quasi-streamwise vortices, which will be first identified and extracted from the dataset, then ensemble-averaged and eventually compared across the various control techniques.

%--------------------------------------------
\subsection{Extracting and averaging the QSV}
\label{sec:extraction}

The extraction procedure is based on the swirling strength criterion, introduced by Zhou et al. \cite{zhou-etal-1999}, and resembles an equivalent procedure illustrated by Jeong et al. \cite{jeong-etal-1997} based on the $\lambda_2$ criterion. The swirling strength criterion exploits the fact that, when the velocity field exhibits locally spiralling streamlines, the velocity gradient tensor $\vect{\nabla u}$ possesses a pair of complex conjugate eigenvalues: the swirling strength $\lambda_{ci}$ is the absolute value of their imaginary part. According to \cite{zhou-etal-1999}, vortex cores are detected where $\lambda_{ci}$ exceeds a threshold value $\lambda_{ci}^{th}$. The main advantage of the $\lambda_{ci}$ criterion over the $\lambda_2$ criterion is that directional information is available through the eigenvectors of $\vect{\nabla u}$. In particular, the eigenvector associated to the real eigenvalue lies along the direction in which the vortex is stretched or compressed; hence its direction, evaluated at the center point of a vortex (defined as the point where $\lambda_{ci}$ is maximum) defines the vortex axis. To compare the OW cases with YHK14, where the $\lambda_2$ criterion was used with a threshold value of $\lambda_2^{th}=0.02$, we started from the relation $\lambda_{ci}^{th}=(\lambda_2^{th})^{0.5}=0.141$ (which is only valid when the eigenvalues are real, \cite{chakraborty-balachandar-adrian-2005}) and then slightly adjusted $\lambda_{ci}^{th}$ to maximize the percentage of structures that are identically extracted by the two criteria, ending up with $97.6\%$ of coincident structures extracted with $\lambda_2^{th}=0.02$ and $\lambda_{ci}^{th}=0.145$.

\begin{figure}
\centering
\includegraphics[width = 0.75\textwidth]{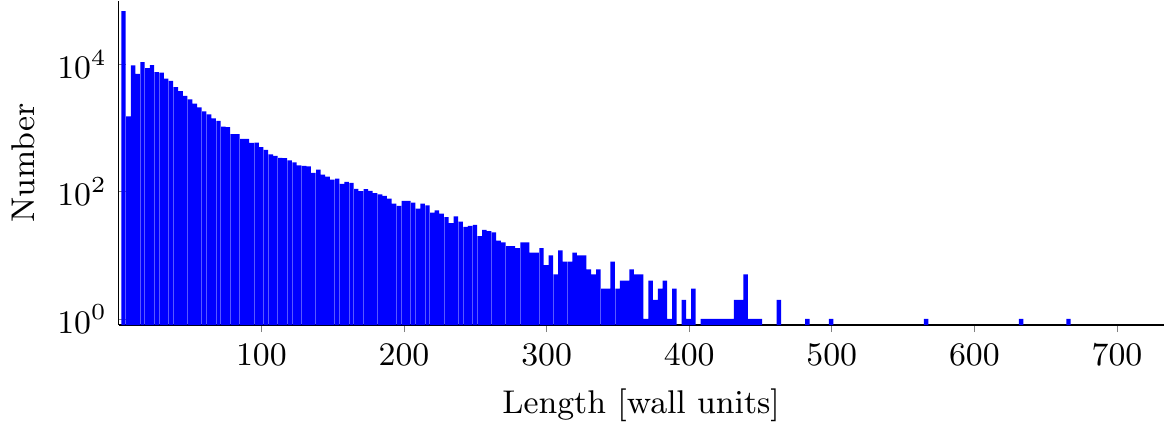}
\caption{Statistical distribution of the length of the extracted vortex candidates (reference case).}
\label{fig:lengthpdf}
\end{figure}

Once applied to the whole dataset, the extraction procedure yields a set of spatially oriented candidate structures with different lengths. The length of a structure is defined here as the length of the segment formed by the intersection of the vortex axis with the smallest box, with edges parallel to the $x,y,z$ directions, containing all the points of the structure. An example of the length distribution across the population of extracted QSV candidates is shown in figure \ref{fig:lengthpdf}. Several vortex candidates of extremely short length are visible, with the smallest length bin sowing a significant peak. This is a known problem, and the original Ref. \cite{jeong-etal-1997} proposed a threshold length to eliminate excessively short structures. In this work, a vortex candidate is discarded when shorter than 50 wall units. This threshold value is comparatively rather low: for example Jeong et al.\cite{jeong-etal-1997} excluded vortices shorter than 150 wall units. In the reference case, our threshold is such that only $13.4\%$ of the vortex candidates are retained. In the controlled cases, we opted for discarding the same percentage of candidate vortices considered in the reference case, to avoid the assumption that control does not affect the length of QSV. Alternatively, the threshold length could have been kept fixed at 50 wall units, thus discarding a different percentage of vortices for each controlled case. The two criteria are conceptually different, but they have been checked to produce very similar outcomes: keeping fixed the percentage of retained structure results in a threshold length that varies from $49.5$ (OW1) to $47.1$ (TW2) wall units, supporting a minor influence of the control on the length of the QSV. 

\begin{figure}
\centering
\includegraphics[]{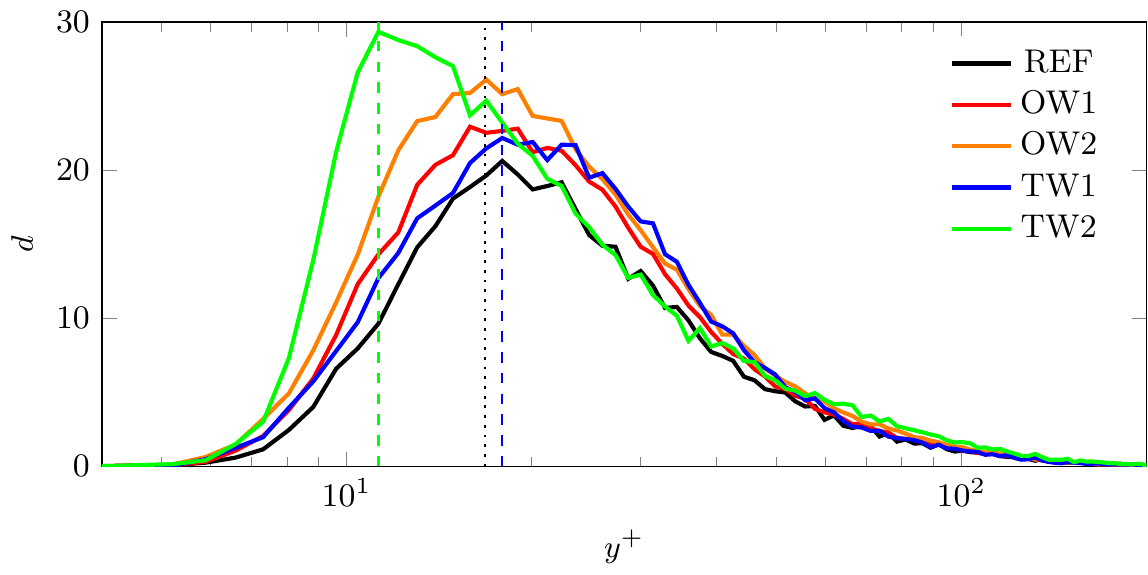}
\caption{Wall-normal distribution of the density $d$ of educed vortices. The black dotted line at $y^+=16.8$ marks the position of the maxima determined by YHK14 for OW, whereas the colored dashed lines mark the maximum for TW1 and TW2.}
\label{fig:ydistr}
\end{figure}

For the five datasets, the density $d$ per unit volume of retained QSV is plotted in Fig. \ref{fig:ydistr} as a function of the wall-normal position. The quantity $d$ is defined, at each grid point $y_i$ in the wall-normal direction, as the total number of extracted vortices divided by the number of flow fields and by the slice volume $L_x \times L_z \times (y_{i+1}-y_{i-1})/2$, so that the integral of the curves in fig.\ref{fig:ydistr} multiplied by the volume of the computational domain gives the total number of extracted structures.

It is expected \cite{ibrahim-etal-2021} that changes in friction drag reflect into the wall-normal position of turbulent structures. Indeed, OW1 and OW2 curves are qualitatively similar to the reference case, and the maximum of the distributions consistently occurs near the position $y^+ = 16.8$ identified by YHK14. The three-fold increase in the number of detected events reported in YHK14 for a case very similar to OW2 is not confirmed by the present data, which show a more modest increment. It should be noticed, though, that YHK14 does not mention the use of a threshold length in their eduction procedure. For the StTW cases, the wall-normal distribution changes more evidently: the drag-reducing case TW1 shows a slight shift of the peak further from the wall at $y^+=17.9$, whereas the drag-increasing TW2 case has the peak significantly shifted towards the wall at $y^+=11.3$. 

To create an ensemble average of the extracted QSV, we follow the procedure described by YHK14. The educed structures are centered around the wall-normal position $y_c$ of the maximum of the QSV density $d(y)$ (from here on, a subscript or superscript $c$ will identify conditionally-averaged quantities): all the events located between the two grid points above and below $y_c$ are identified, and the surrounding flow fields are horizontally shifted to align at the center. In the subsequent averaging procedure, to maximize the statistical sample every available vortex is considered: this involves putting together structures rotating in both directions, as well as structures located near both walls. The sense of rotation is computed as the scalar product of the random vorticity fluctuation vector and the real eigenvector of $\vect{\nabla u}$ in the vortex center. Negatively-rotating vortices are then flipped about the $x-y$ plane with respect to the structure center. QSV belonging to the opposite wall are flipped about the channel midplane at $y=h$, and, in case of positive rotation, flipped once more about the $x-y$ plane. Moreover, in the cases with wall forcing, the phase of the forcing is also shifted by $\pi$ for events with negative rotation. For the cases with less structures entering the conditional average (TW1 and TW2), approximately 1200 QSVs are considered for each period. This is sufficient to produce well resolved structures and, for each realization, it is in line with previous works \cite{jeong-etal-1997}.  

%-----------------------------------------------------
\subsection{Changes of the conditionally-averaged QSV}
\label{subs:QSVmod}

\begin{figure}
\centering
\includegraphics[width=0.75\textwidth]{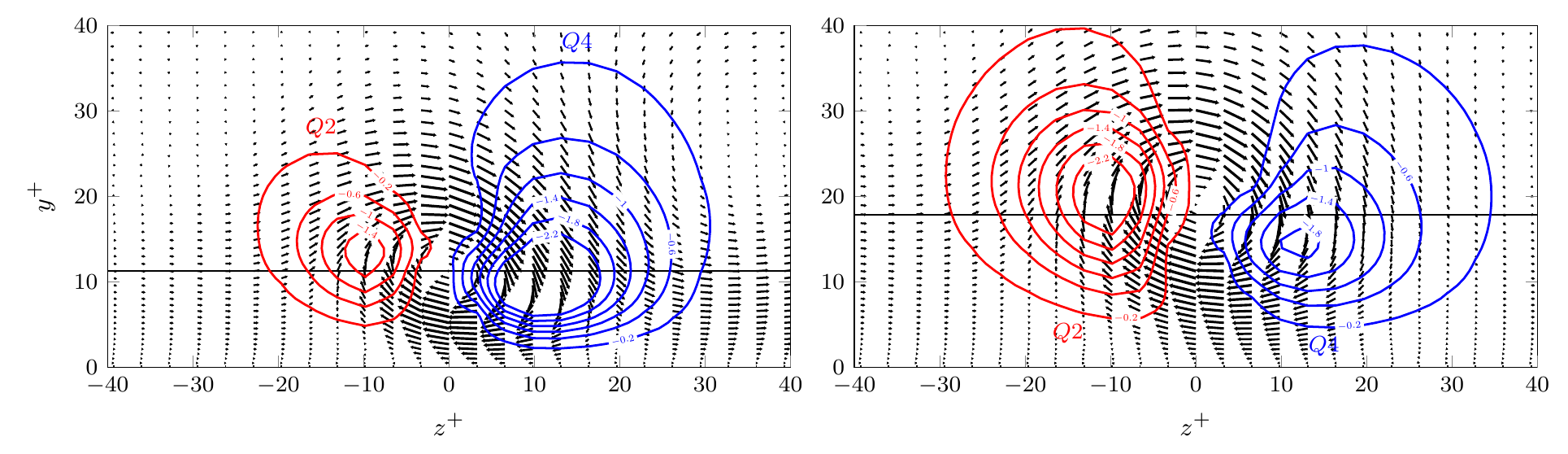}
\caption{Conditionally-averaged flow field for the reference case; cross-stream velocity vectors $(v_c,w_c)$ and contributions $Q_2$ (red lines) and $Q_4$ (blue lines) in the midplane at $x=0$. QSV extracted at the reference wall distance $y^+=11.3$ (left) and $y^+=17.9$ (right), shown by the horizontal lines. Q2 and Q4 isocontours range from $-0.2$ to $-2.2$ with a $-0.4$ step.}
\label{fig:condav_ref}
\end{figure}

The conditionally-averaged QSV is considered first for the reference case in figure \ref{fig:condav_ref}. It is computed and plotted twice, namely for structures extracted at the two wall distances corresponding to $y_c^+=11.3$ and $y_c^+=17.9$. This facilitates the comparison with the controlled cases, where (see Fig. \ref{fig:ydistr} above) the vortex density for TW1 (drag reduction) peaks at $y^+=17.9$ and that for TW2 (drag increase) peaks at $y^+=11.3$. Comparing the two averaged velocity fields in the reference case shows that the QSV features, albeit qualitatively similar, do change with their wall distance. In particular, the $Q_2$ contribution is more intense further form the wall, whereas $Q_4$ is strengthened nearer to the wall.

%\begin{figure}
%\centering
%\includegraphics[]{fig/condav_OW1.pdf}
%\caption{Conditionally-averaged flow field for case OW1, at eight phases along the oscillation cycle, with QSV extracted at $y_c^+=16.9$. Lines and vectors as in figure \ref{fig:condav_ref}. The additional right panel plots the phase-averaged mean velocity profile.}
%\label{fig:condav_OW1}
%\end{figure}

%\begin{figure}
%\includegraphics[]{fig/condav_OW2.pdf}
%\caption{Same as figure \ref{fig:condav_OW1}, for case OW2, with QSV extracted at $y_c^+=16.9$.}
%\label{fig:condav_OW2}
%\end{figure}

\begin{figure}
\includegraphics[width=\textwidth]{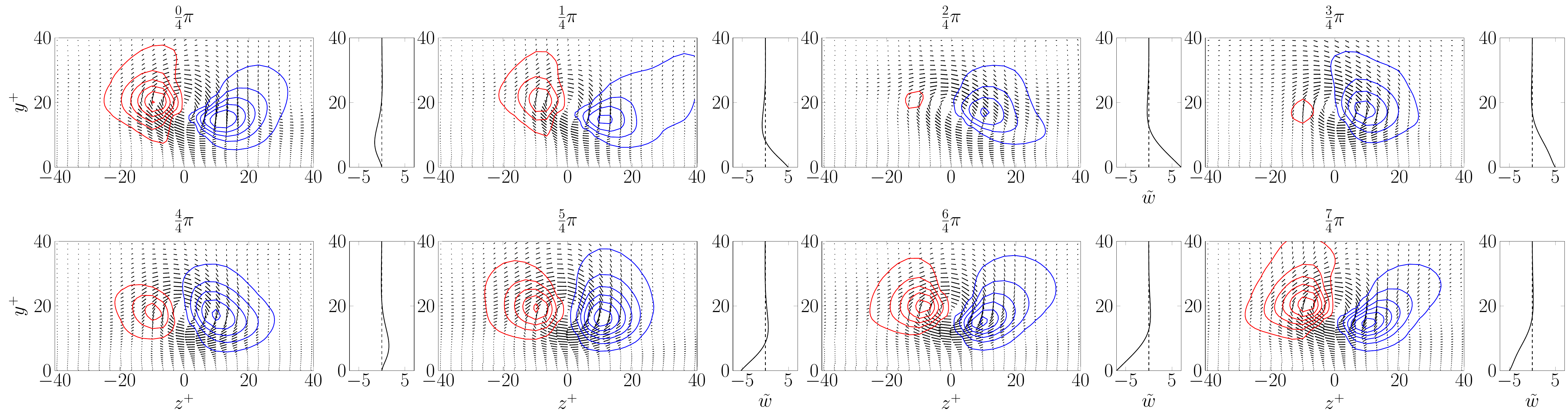}
\caption{Conditionally-averaged flow field for case TW1, at eight phases along the oscillation cycle, with QSV extracted at $y_c^+=17.9$. Lines and vectors as in figure \ref{fig:condav_ref}. The additional right panels plot the phase-averaged mean velocity profile.}
\label{fig:condav_TW1}
\end{figure}

\begin{figure}
\includegraphics[width=\textwidth]{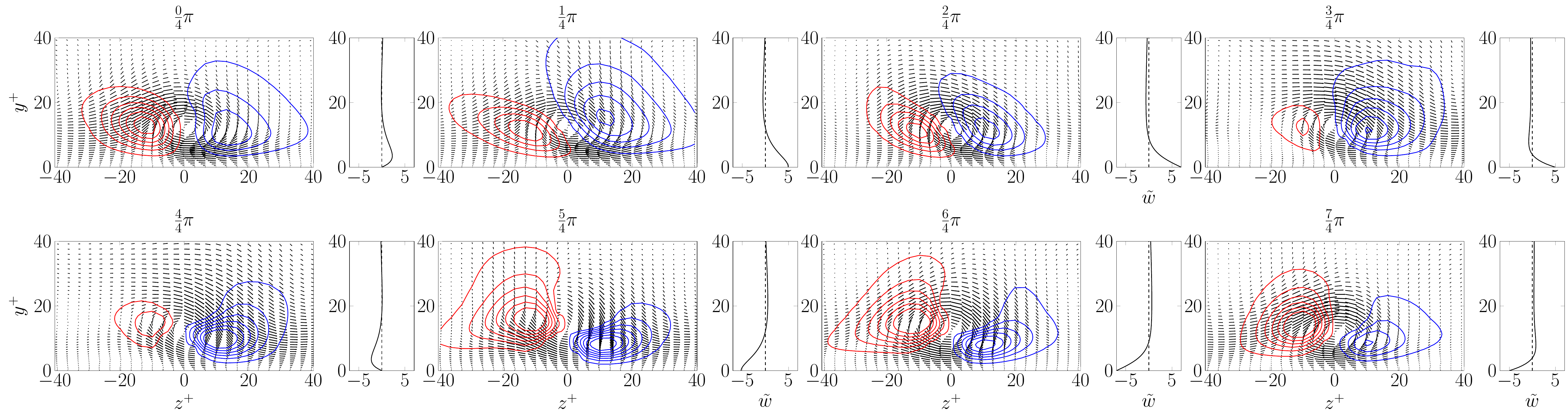}
\caption{Conditionally-averaged flow field for case TW2, at eight phases along the oscillation cycle, with QSV extracted at $y_c^+=11.3$. Lines and vectors as in figure \ref{fig:condav_ref}. The additional right panels plot the phase-averaged mean velocity profile.}
\label{fig:condav_TW2}
\end{figure}

The conditionally-averaged flow fields for cases OW1 and OW2 (not shown for brevity) qualitatively confirm the picture drawn by YHK14, with the $Q_2$ and $Q_4$ contributions possessing an evident phase dependency, in terms of both position and strength. Figures \ref{fig:condav_TW1} and \ref{fig:condav_TW2} plot the conditionally-averaged flow field for cases TW1 and TW2: QSV are extracted at the position $y_c^+$ where the corresponding density has a peak, at eight equally spaced phases along the forcing period.
As for OW, drag reduction brings about a suppression of Q2 events, at least for TW1. The minimum Q2 contour levels for TW1 appear at a phase which is the same of OW1 (but not OW2). The drag-increasing TW2 case, however, shows a remarkable range of variation in $Q_2$ values along the cycle, varying from a minimum at $\phi = 3/4 \pi$ to a significantly larger maximum at $5/4 \pi$ and $6/4 \pi$. Changes of Q4 events are of lesser evidence, as expected from Fig. \ref{fig:hist}, but generally sweeps do appear as more intense in controlled cases than in the uncontrolled one.

These features can be put in direct relation to the changes $\Delta Q_i$ discussed in Fig. \ref{fig:hist}, so that the share of each quadrant modification to the build up of the total drag change expressed by $\Delta U_b$ can be quantified. As in YHK14, this is obtained by integrating over a volume $V_c$ the weighted Reynolds shear stresses due to conditionally-averaged Q2 and Q4 events. The considered $V_c$ is a box extending for $100 \times 80$ wall units in the $x$ and $z$ directions, with its center coincident with the QSV center, and which extends from the wall up to $y^+=64$. The box size is similar to that employed in YHK14, with marginally lower wall-parallel size and larger height. The computed integral value is then multiplied for the vortex density $d$ per unit volume at the $y_c$ location of the averaged QSV to yield a volume-integrated quadrant contribution expressed in wall units:
\begin{equation}
Q^c_i (\phi) = -d \int_{V_c} \left( 1-\frac{y^+}{Re_\tau} \right)
\left( -\overline{u^c v^c}(\phi) \right)_i dV .
\end{equation}

\begin{figure}
\includegraphics[width=0.75\textwidth]{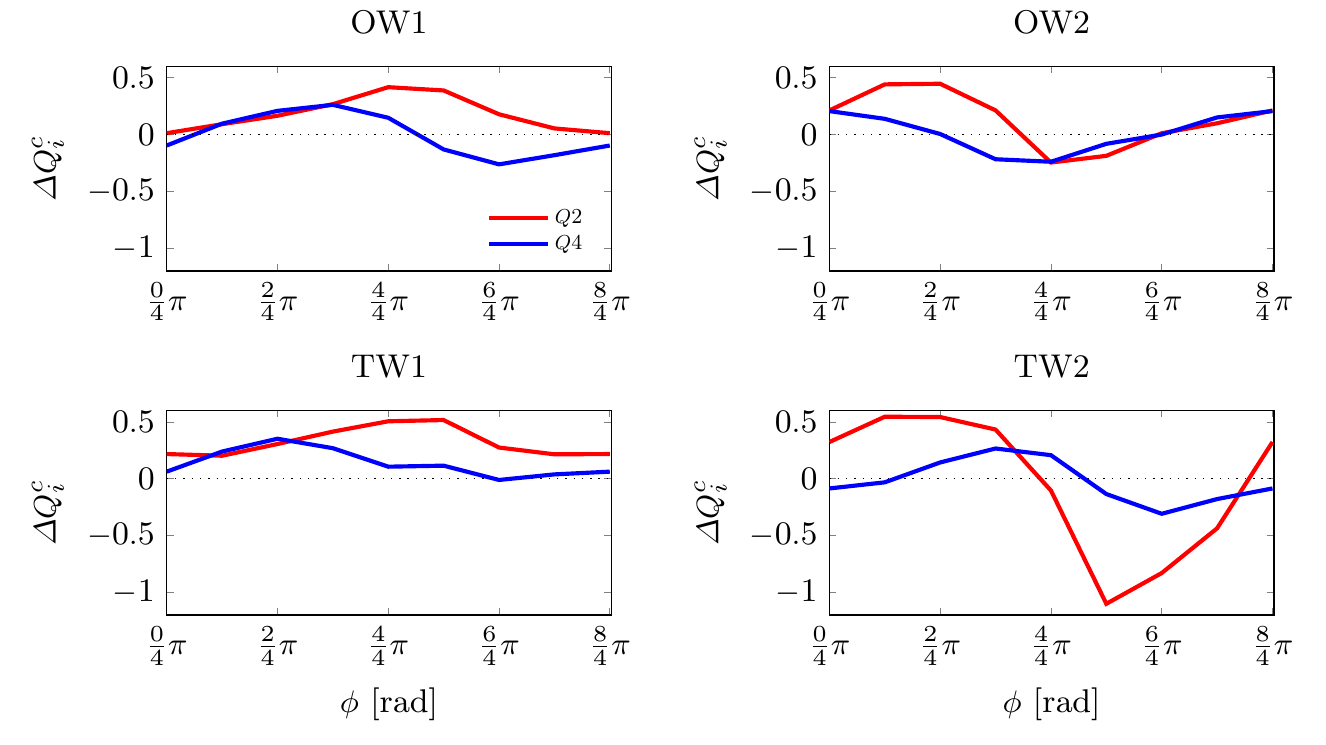}
\caption{Density-multiplied volume integral of the weighted Reynolds shear stresses arising from Q2 (red) and Q4 (blue) events: difference with the reference case against phase. }
\label{fig:Q2Q4int}
\end{figure}

Changes from the baseline case can thus be isolated, complementing the simple total change $\Delta Q_i$ with the more informative quantity $\Delta Q^c_i(\phi)$ that keeps track of the forcing phase. In Fig. \ref{fig:condav_ref}, it was appreciated how the features of the extracted QSV depend upon the wall-normal position $y_c^+$ of the extraction. For this reason, the contributions $Q^c_i$ are computed at a few nearby grid points $y^+_c$ centered around the wall-normal peak of the vortex distribution, and then summed together. The results are eventually plotted in Fig. \ref{fig:Q2Q4int} as a function of the phase $\phi$ of the forcing for all the 4 considered cases. The integral along the phase of the changes $\Delta Q_i^c(\phi)$ corresponds to the global change $\Delta Q_i$ discussed in Fig.\ref{fig:hist}. The present results confirm the involvement of the quasi-streamwise vortices in the drag reduction process for both the oscillating wall and the streamwise-traveling waves.

\begin{figure}
\includegraphics[width=0.5\textwidth]{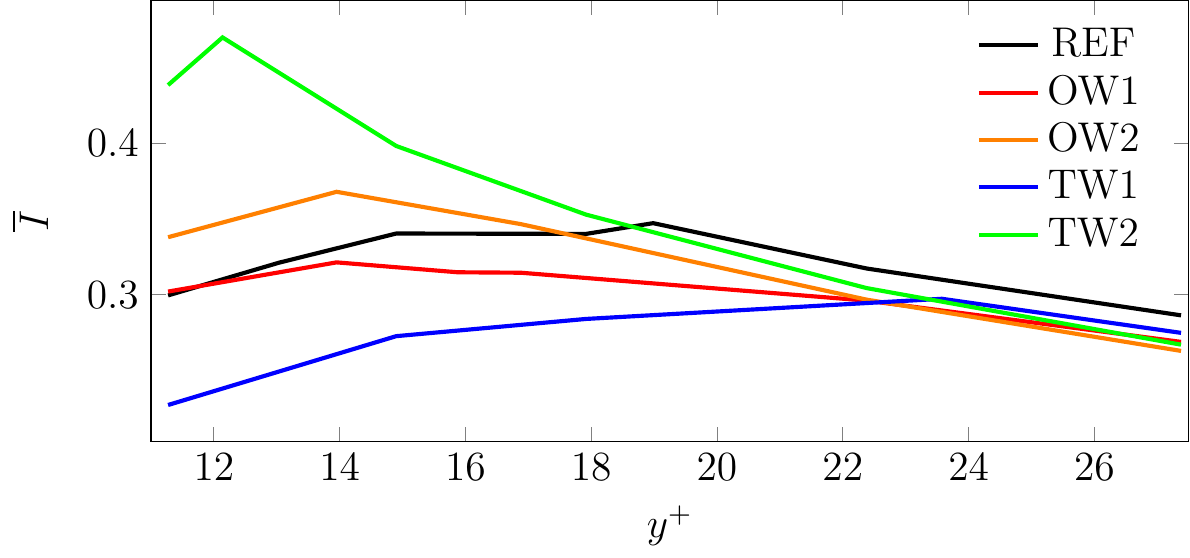}
\caption{Wall-normal distribution of the phase-averaged vortex intensity $\overline{I}$ of conditionally-averaged QSV: comparison between controlled and uncontrolled cases.}
\label{fig:intensity}
\end{figure}

Both Figures \ref{fig:condav_TW2} and \ref{fig:Q2Q4int} indicate large phase shifts of $Q_2$ and $Q_4$ events for TW2. To investigate their origin, we considered the intensity of the conditionally-averaged structures. The intensity $I(y^+,\phi)$ of a QSV has been defined as the integral over the vortex core of the streamwise component of the vorticity $\omega_x$. The vortex core has been identified, in this application, as the region where $\omega_x(y^+,\phi)_x > 0.5\omega_x(y^+,\phi)_{x,max}$, being $\omega_{x,max}$ the maximum of the streamwise vorticity at the phase $\phi$ for QSV with center in $y^+$. 
The value of the phase-averaged intensity $\overline{I}(y^+)$ as a function of $y^+$ is represented in Fig. \ref{fig:intensity} for the different control strategies. In presence of drag increase (TW2) or moderate drag reduction (OW2), the structures close to the wall are intensified with respect to the reference case; the opposite happens when drag reduction is large (TW1, OW1). The distance  from the $\overline{I}_{REF}$ curve becomes larger for larger drag changes. 
When $y^+$ increases, all the controlled vortices are damped: TW2 and OW2 have a decreasing trend while OW1 and TW1 tend towards the value of $\overline{I}_{REF}$. By comparing Fig. \ref{fig:ydistr} and \ref{fig:intensity}, it can be noticed that the maxima of $\overline{I}$ and those of the QSV's wall-normal distribution for TW2 are very close (despite not coincident): $Q_2$ and $Q_4$ suppression and enhancement described in YHK14 affect intense and numerous structures producing the large phase shifts observed in Fig. \ref{fig:condav_TW2} and \ref{fig:Q2Q4int}.
%%NOTA: it would be interesting also to observe negative \omega_x

It was mentioned above for OW that $Q_2$ and $Q_4$ conditional fields change with the phase, but their wall-normal location, identified by the position of their minima, remains constant. A peculiar feature of the TW cases, visible in figures in Fig.\ref{fig:condav_TW1} and \ref{fig:condav_TW2}, is thus the evident vertical movement of $Q_2$ and $Q_4$ conditional fields around the vortex center during the forcing cycle, with the center of the $Q_2$- and $Q_4$-structures bouncing vertically. This phase-dependent wall-normal movement of the $Q_2$- and $Q_4$-structures is more pronounced for the TW2 case, and seems to be related to an enhancement of the Reynolds-stress events and hence to drag increase. In fact, the phases $5/4 \pi$ nd $6/4\pi$, identified above to show the strongest Q2 and Q4 events, also exhibit the largest positive (negative) wall-normal shift of the Q2 (Q4) structure. 
It is thus natural to link this shift to the changes to the phase-averaged strain-rate tensor $\aver{S}_{ij}$ induced by the streamwise dependence of the phase-averaged velocity field: indeed $\partial \aver{u} / \partial x$, $\partial \aver{v} / \partial x$ and (to a larger extent, especially near the wall) $\partial \aver{w} / \partial x$ are non-zero for the StTW cases, where $\phi$ contains a dependence on $x$.

\begin{figure}
\centering
\includegraphics[width=0.5\textwidth]{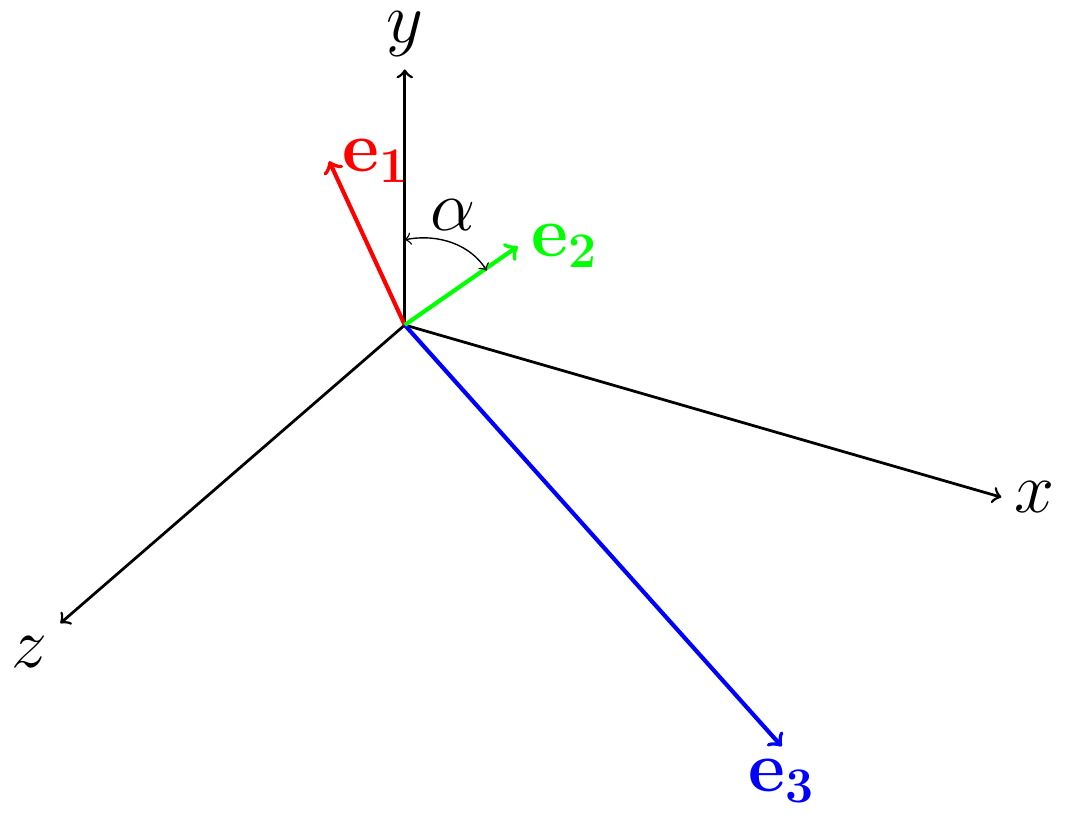}
\caption{Sketch of the stress state for StTW forcing. $\alpha$ is the angle between the intermediate eigenvector of the strain-rate tensor and the wall-normal axis; for OW $\alpha$ is fixed to $\alpha = \pi/2$.}
\label{fig:strainrep}
\end{figure}

For the OW cases, the second eigenvalue $\lambda_2$ of $\aver{S}_{ij}$ is zero, and the principal stresses correspond to the other two non-zero eigenvalues:
\begin{equation}
\lambda_{1,3} = \pm \sqrt{ \left( \frac{\partial \aver{u}}{\partial y} \right)^2 + \left( \frac{\partial \aver{w}}{\partial y} \right)^2 }.
\end{equation}

To obtain the principal stresses, $\lambda_{1,3}$ are multiplied by the dynamic viscosity $\mu$ and by the associated normalized eigenvector, yielding the two non-zero eigenvectors:
\[
\vect{e_{1,3}}(y) = \frac{\mu}{\sqrt{2}}
\begin{Bmatrix}
\displaystyle
\frac{\partial \aver{u}}{\partial y} \\
\displaystyle
\lambda_{1,3}\\
\displaystyle
\frac{\partial \aver{w}}{\partial y} 
\end{Bmatrix} .
\]

The stress state is thus planar, with the resultant vector aligned to the $y$ axis; moreover, the spatial orientation of the vector changes with the phase, at any wall-normal position. In the StTW cases, however, all the three eigenvalues are non-zero, and the stress state has a three-component resultant, with an inclination to the $y$ axis that is time-dependent, expressed by the angle $\alpha$ between the direction of the eigenvector $\mathbf{e_2}$ associated to the second largest eigenvalue $\lambda_2$ and the $y$ axis. 

\begin{figure}
\includegraphics[width=0.75\textwidth]{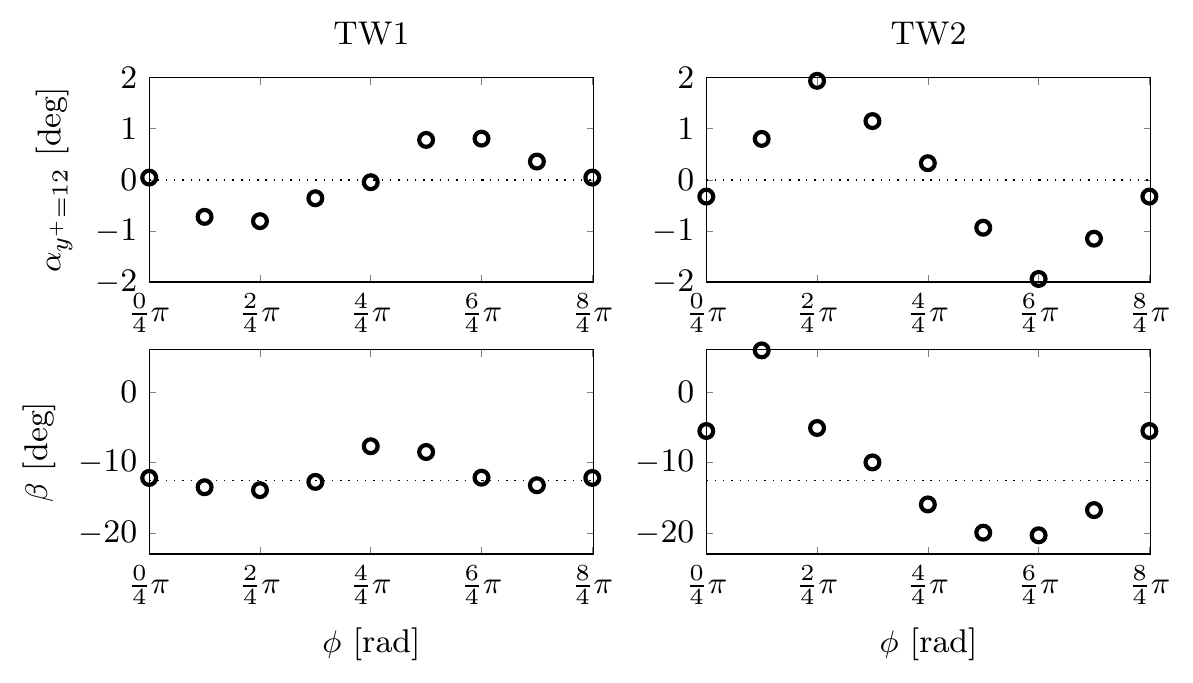}
\caption{Comparison between the shear angle $\alpha$ at $y^+=12$ with the angle $\beta$ between the line connecting $Q_2$ and $Q_4$ maxima at $\Delta x = 0 $ and the $z$ axis. Dotted lines indicate the reference values.}
\label{fig:Vortexroll}
\end{figure}

To assess whether the shear angle $\alpha$ is linked to the conditionally extracted structures, especially in the so far unexplored StTW case, figure \ref{fig:Vortexroll} compares $\alpha$ computed at $y^+=12$ with the angle $\beta$ formed by the straight line connecting the maxima of the $Q_2$ and $Q_4$ fields at $\Delta x=0$ and the $z$ axis. The specific wall distance $y^+=12$ is chosen after Duque-Daza et al.\cite{duque-etal-2012} who found it to be the likely location of the most dynamically significant velocity streaks. The two angles vary with the phase in a very similar way, confirming the dynamical link between the shear at $y^+=12$ and the vertical bouncing of the conditionally averaged fields. Moreover, we observe that both $\alpha$ and $\beta$ are larger in absolute value for the TW2 case, supporting the idea that large wall-normal excursion of the Reynolds shear stress structures relate to drag increase.
%The shear angle in figure \ref{fig:Vortexroll} has been computed analytically (by using a model expression for the mean streamwise velocity profile to compute $\partial \aver{u} / \partial y$, and the analytical GSL solution to compute $\partial \aver{w} / \partial y$, with the additional assumption that $\partial \tilde{v}/\partial x = 0$ is negligible), but the next figure \ref{fig:vc} confirms that predicting or measuring the shear angle does not matter much, provided the forcing yields drag reduction and the laminar GSL solution remains predictive of the spanwise flow. 

%NOTA: y+=12 is recurrent. Maybe: QSV birth is the result  of linear amplification mechanism, happening at a wall normal position scaling with viscous units and directly influenced by Stokes layer. The modifications can be reflected non-linearly in the vortex behavior.

%\begin{figure}
%\center
%\includegraphics[]{./fig/TW2detail.pdf}
%\caption{Conditionally averaged cross-stream velocity field at $\Delta x^+=0$, with isocontours for the wall-normal velocity component at $\phi=3/4\pi$ (left, moderate $\beta$) and $\phi=6/4\pi$ (right, large $\beta$) for the TW2 case.}
%\label{fig:vc}
%\end{figure}

%Figure \ref{fig:vc} plots isocontours for the wall-normal velocity component of the conditionally averaged structure for the drag-increasing TW2 case, at phases corresponding to a moderate and large $\beta$, namely $\phi=3/4\pi, 6/4\pi$: at the latter phase the vertical component is larger, contributing to the enhancement of Reynolds stresses and consequently to the increase of drag. 

\begin{figure}
\includegraphics[width=\textwidth]{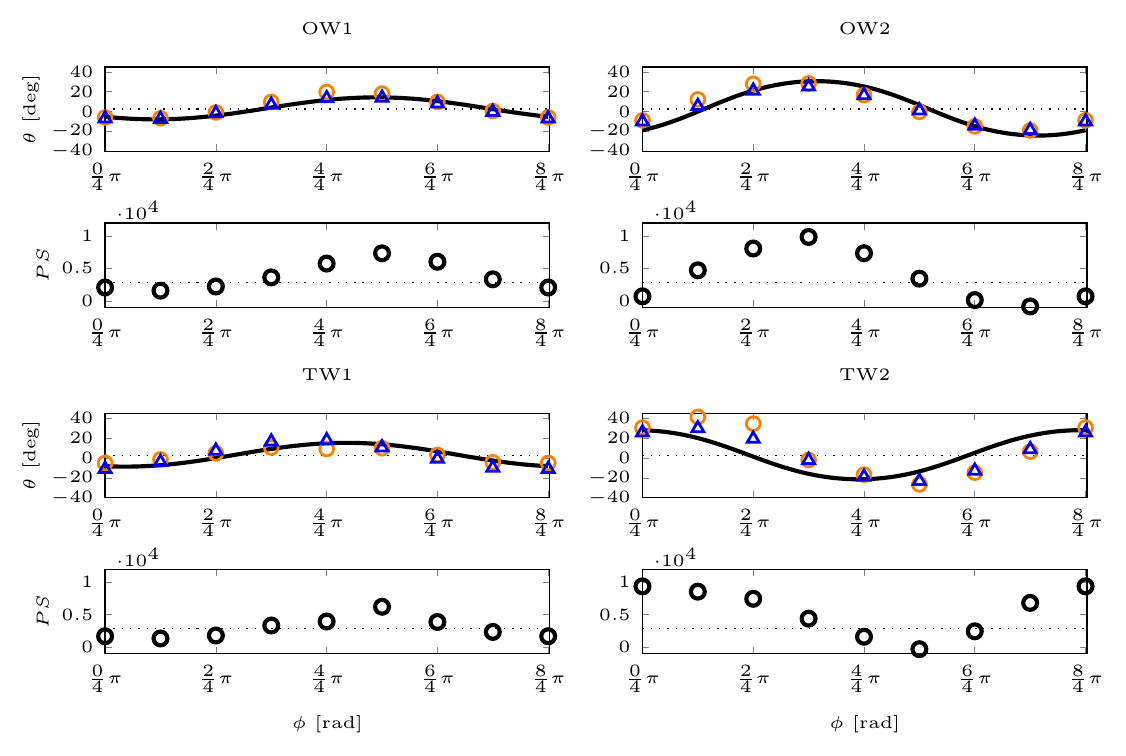}
\caption{Tilt angle $\theta$ and volume integral of the pressure-strain term vs phase. The continuous line is the prediction based on the laminar GSL solution, whereas circles are computed from the actual spanwise Stokes layer; triangles are actual measurements from conditional average. The horizontal dotted line corresponds to the reference value.}
\label{fig:Angvsphase}
\end{figure}

Owing to the streamwise shear, the conditionally averaged structures possess a non-zero tilt angle $\theta$, the angle between the direction of the vortex axis (as defined in Subsection \ref{sec:extraction}) and the streamwise direction. For  the uncontrolled case, Jeong at al.\cite{jeong-etal-1997} computed $\theta_0=4^\circ$, in good agreement with $\theta_0=2.5^\circ$ obtained here, whereas $\theta_0=9^\circ$ is reported by YHK14. The spanwise forcing acts on the vortex and consequently modifies the tilt angle. According to YHK14, the phase-varying tilt angle $\theta(\phi)$ can be effectively estimated as:
\begin{equation}
\theta = -\arctan \left( \frac{\partial \tilde{w}/\partial y}{\partial \bar{u}/\partial y} \right) + \theta_0,
\label{eq:theta}
\end{equation}
where at the denominator $\partial \bar{u}/\partial y$ is replaced by $\partial \tilde{u} / \partial y$ since the periodic fluctuation of the streamwise shear is negligible. Moreover, in Ref.~\cite{jeong-etal-1997} it is pointed out that the pressure-strain term $-p \partial u / \partial y$ is related to the tilt angle, as a highly tilted vortex generates a large negative pressure strain, transferring energy from the streamwise velocity component towards spanwise and wall-normal ones, thus increasing the rotational motion in the $y-z$ plane. 

In Fig.~\ref{fig:Angvsphase} the tilt angle measured for the conditionally averaged vortex is presented, and compared to the shear angle computed from the laminar GSL solution \cite{quadrio-ricco-2011}, as well as to the actual spanwise Stokes layer profile measured from DNS data, to fully probe Eq.\ref{eq:theta}. In YHK14 the shear angle was evaluated at $y^+=15$, the rationale being that the peak in the wall-normal distribution of vortices for their OW cases was found at $y^+=16.8$, and that $\theta$ should be affected by the shear immediately below the peak, since the QSV has a leg that extends downwards towards the wall.
Following the same rationale, data in figure \ref{fig:Angvsphase} are computed for structures extracted at positions that, for each case, are just below the respective distribution peak, namely $y^+=15$ for OW1 and OW2, $y^+=16$ for TW1 and $y^+=9$ for TW2. Except the TW2 case, where drag increase implies a difference between the laminar GSL solution and the actual GSL profile, the three different evaluations of $\theta$ are nearly coincident. Comparing Fig.~\ref{fig:Angvsphase} and Fig.~\ref{fig:condav_TW2}, it is observed that when the QSVs are aligned along the streamwise direction, with $\theta$ passing from negative to positive values, Q2 is drastically reduced. YHK14 found the same for OW, attributing it to the counteraction of the shear stress $\partial \tilde{w}/\partial y$ on the vortical motion under the vortex center. Half a cycle later, $\theta$ is still close to zero but the shear stress and the velocity induced by the QSV are aligned, so that the suppressing effect vanishes. Furthermore, Fig.~\ref{fig:Angvsphase} plots the volume integral $PS$ of the pressure-strain term, i.e.
\[
PS = \int_{V_c} - p_c \frac{\partial u_c}{\partial x} dV .
\]

Its direct relationship with the instantaneous tilt angle is confirmed for the OW cases, and shown here to extend to the StTW cases, with a large pressure strain for TW2 indicating the production of $y-z$ fluctuations.

\section{Scaling of drag reduction}
\label{sec:dragchanges}

In this section, we attempt to relate the physical properties discussed in the previous section with the drag changes induced by the travelling waves. The main goal is not to propose yet another empirical relationship for predicting drag reduction. That could be more accurately accomplished by interpolating the abundant high-fidelity DNS data available in the literature (see, for instance, \cite{gatti-quadrio-2016}), where the majority of the parameter space has already been spanned. Instead, by inspecting relationships that describe well the drag reduction of travelling waves, we aim to highlight which physical parameters are related to drag reduction (or to the lack thereof). 

As a starting point, the correlations proposed in the literature are considered. These are mostly restricted to the OW case. While they can be easily extended to the TW cases, we show that a trivial extension does not yield satisfactory results.  
\begin{figure}
\centering
\includegraphics[]{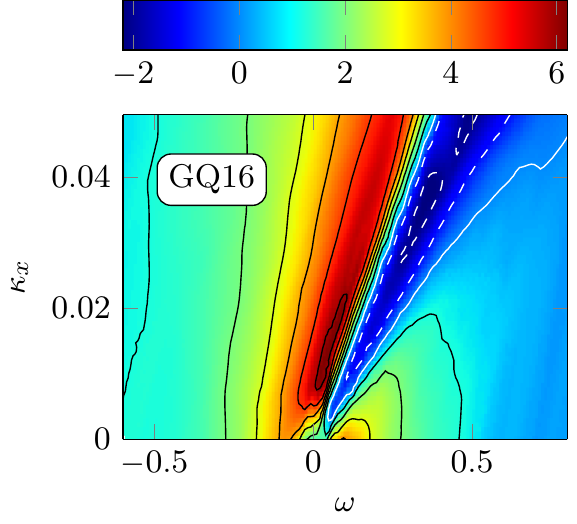}
\includegraphics[]{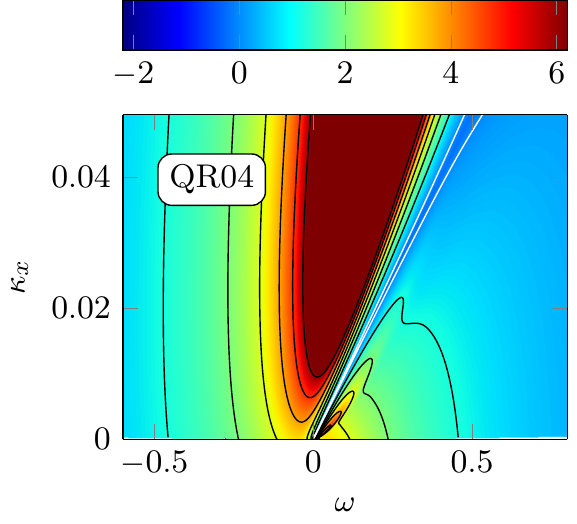}\\
\includegraphics[]{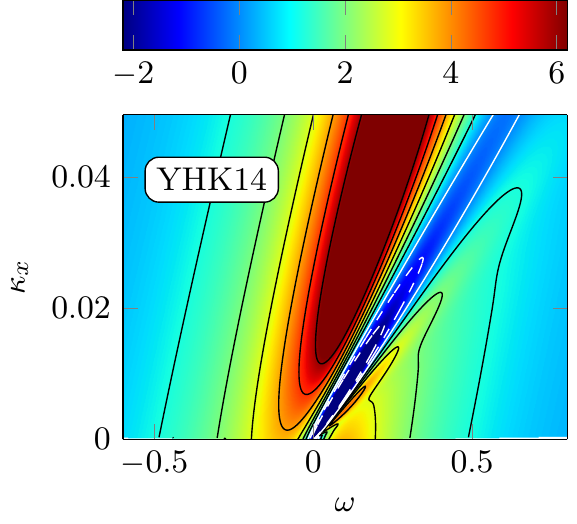}
\includegraphics[]{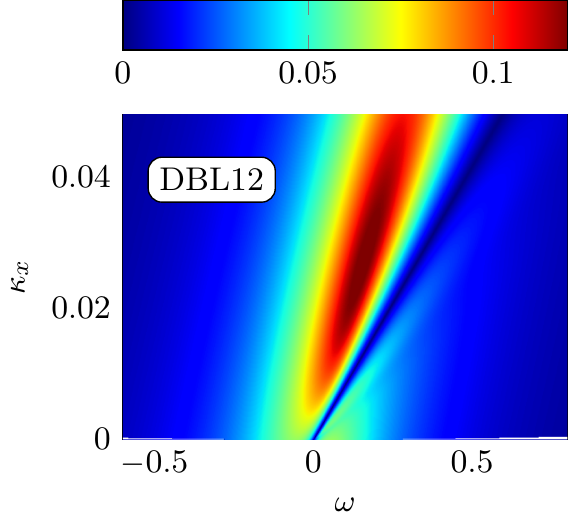}
\caption{Top left: reference drag reduction data at $A=12$ from Gatti \& Quadrio \cite{gatti-quadrio-2016} expressed in terms of $\Delta U$. The other panels plot predictions of drag reduction with formulas available for OW and extended to TW: top right from Quadrio \& Ricco \cite{quadrio-ricco-2004}, bottom left from YHK14 \cite{yakeno-hasegawa-kasagi-2014}. The bottom right panel, derived from Duque-Daza et al. \cite{duque-etal-2012}, does not allow a quantitative prediction, hence its color scale does not correspond to previous figures: the map is for the quantity $\partial^2 \tilde{w} /\partial y^2 |^{rms}$ at $y^+=12$.}
\label{fig:corrs2}
\end{figure}
The reference dataset of choice is the publicly available one by Gatti \& Quadrio \cite{gatti-quadrio-2016}; we extract from it data computed at $Re_\tau=200$ and $A=12$. The drag reduction as a function of the remaining control parameters is plotted in the upper left panel of Figure \ref{fig:corrs2} on the $\omega - \kappa$ plane, where drag reduction is expressed in terms of the upward shift $\Delta U$ of the mean velocity profile, which is the correct, Reynolds-independent representation. The conversion between the drag reduction rate $R$ and $\Delta U$ is performed with equation (4.8) of \cite{gatti-quadrio-2016}, which we also report here for convenience:
\begin{equation}
    \Delta U = U_{b,0} \left[(1-R)^{-1/2} - 1 \right] \, ,
\end{equation}
assuming a reference uncontrolled channel at $Re_\tau=200$, i.e. with $U_{b,0}=15.9$ \cite{gatti-quadrio-2016}. 

The remaining three panels plot drag-reduction relationships based on available or easily extended formulas. The top right panel shows the extension to TW of the empirical correlation 
\begin{equation}
    R = 1.31 S - 0.27
    \label{eq:Scorr}
\end{equation}
proposed by \cite{quadrio-ricco-2004} and recast in terms of $\Delta U$ as described above. Eq.\ref{eq:Scorr} contains the parameter $S$, first introduced by J.-I. Choi \emph{et al.} \cite{choi-xu-sung-2002} to empirically correlate drag reduction from OW with properties of the Stokes layer. The dimensionless parameter $S$ is defined as 
\[
S = a_m \frac{\ell}{A} ,
\]
where $a_m$ is the maximum spanwise acceleration of the Stokes layer during the oscillation cycle, and $\ell$ is a length-scale related to the thickness of the Stokes layer, which can be interpreted as its penetration depth into the bulk of the flow. Both $a_m$ and $\ell$ contain further user-defined parameters. First, the wall-normal position at which $a_m$ is evaluated needs to be specified. Then, in the definition of $\ell$, a conventional velocity threshold $W_{th}$ appears, representing the maximum velocity required to define a nominal Stokes layer thickness. In Ref.\cite{choi-xu-sung-2002} $a_m$ is evaluated at $y^+=5$, and $W_{th}=0.5$ is chosen based on physical considerations. The $S$ parameter was later elaborated and improved in Ref.\cite{quadrio-ricco-2004}, where it was shown that evaluating $a_m$ at $y^+=6.3$ and choosing $W_{th}=1.2$ fits best the drag-reduction data for OW with $T \leq 150$.   

The lower right panel of Figure \ref{fig:corrs2} is not, strictly speaking, a prediction of drag reduction: it shows the term  $\partial ^2 \tilde{w} /\partial y^2 |^{rms}$ evaluated at $y^+ = 12$. The plot is based on the study of Duque-Daza {\em et al.} \cite{duque-etal-2012}, who studied streak amplification in a turbulent channel with the linearized Navier--Stokes equations. They found that the map of the percentage change in streak amplification resembles the drag reduction map. The term $\partial ^2 \tilde{w} /\partial y^2$ enters the linear equation for the streamwise component of the vorticity vector and has also been observed by \cite{duque-etal-2012} to look like the map of drag reduction. 

Finally, the lower left panel stems from the extension to TW of the empirical formula proposed by YHK14 for OW, namely 
\begin{equation}
\Delta U_b =  a     \frac{\partial \widetilde{w}}{\partial y} \Bigg|_{y^+=10}^{rms} - 
              b     \frac{\partial \widetilde{w}}{\partial y} \Bigg|_{y^+=15}^{rms} .
\label{eq:DRpred}
\end{equation}
The formula contains the root-mean-square value of the spanwise shear at $y^+=10$ and $y^+=15$ linearly combined via two empirical coefficients $a$ and $b$. Of the two wall-normal positions involved, the former is where the QSV interacts with the solid wall and thus induces the maximum spanwise velocity. The latter is the average location of the vortex center, where the tilting angle of the QSV correlates well with the behavior of the $PS$ term. Hence, albeit indirectly, the formula somehow includes the physics related to the suppression of Q2 and the enhancement of Q4 events. 

We point out the link between the empirical formula proposed by YHK14 and the streak-amplification analysis by Duque-Daza {\em et al.} \cite{duque-etal-2012}: the former considers $\partial^2 \tilde{w}/\partial y^2$ at $y^+=12$, while the latter is based on the difference between $\partial \tilde{w}/\partial y$  at $y^+=10$ and $y^+=15$, which can be thought of as proportional to a finite-difference approximation of the second derivative in $y^+=12$. 

From a qualitative standpoint, all proposed correlations, once extended to StTWs, capture some features of the actual drag reduction map, but fail at reproducing its overall qualitative behavior. For instance, the YHK14 correlation does not show the expected wavenumber independence at large $|\omega|$, otherwise visible as vertically aligned contours. Moreover, the predicted $\Delta U$ grows unbounded for increasing value of $A$ instead of showing the correct saturating behavior. Finally, the drag-increasing region does not show the correct triangular shape. Indeed, it has been observed in the previous section that the QSV undergo a different and more complex response to the forcing with TW, owing to the more complex stress state in the generalized Stokes layer. For example, the wall-normal bouncing of the $Q2$ and $Q4$ structures flanking the QSV (measured by the angle $\beta$ discussed in Sec.\ref{sec:conditional}) have no counterpart in the OW case.  
Hence, it is not surprising that empirical drag reduction formulas tuned for OW do not perform well for TW. In the following, we try to identify physically sound quantities that can describe the drag changes induced by TW by reworking and improving the empirical formula (\ref{eq:DRpred}).

The lack of a well-defined triangular region for drag increase is addressed first. In \S\ref{subs:QSVmod} the angle $\beta$, associated with an increment of vertical fluctuations and Reynolds stresses, has been related to drag increase. The phase evolution of $\beta$ was found to follow the one of the shear angle at $y^+=12$. This effect should therefore be included via the shear angle $\alpha$ at $y^+=12$ to improve the prediction of drag increase.

To address the failure at predicting a wavenumber-independent drag reduction at large negative frequencies, we observe that the correlation based upon the $S$-parameter (top right of fig.\ref{fig:corrs2}) shows correct drag reduction contours in that region while the YHK14 correlation does not. Hence, elements from $S$ should be incorporated into the prediction formula. Particularly relevant is the acceleration within the Stokes layer, which plays a distinct role in the TW forcing. For OW, a QSV will experience an acceleration on a time scale dictated by the oscillation frequency $\omega$, since the forcing is applied homogeneously at the wall. For TW, a QSV experiences an unsteady forcing only if its convection velocity $U_w \approx 10$ \cite{kim-hussain-1993,quadrio-luchini-2003}, i.e. the velocity with which a QSV travels downstream, differs from the phase speed $c$ of the travelling wave. When such unsteadiness is lacking, i.e. when $U_c \approx c$, no drag reduction is expected. Therefore, the acceleration computed in a reference frame moving with $U_c$, as originally done in Ref. \cite{quadrio-ricco-viotti-2009}, should be included in the empirical correlation for drag reduction.  To this aim, the analytical expression for the GSL velocity profile used to compute the acceleration takes into account the equivalent phase speed $c_{eq} = c - U_w$ by substituting $\omega$ with the equivalent frequency $\omega_{eq} = U_w \kappa_x - \omega$.

To consider the control effect on the entire population of turbulent structures, a modified parameter $S'=a' \ell / A$ is now introduced. While the penetration depth $\ell$ is computed as in the original definition, the acceleration $a'$ is the maximum over the forcing cycle of the GSL acceleration computed after averaging over its penetration depth:
\begin{equation}
a'(\omega, \kappa_x) = \frac{1}{\ell}\int_0^{\ell} a_m (\omega_{eq}, \kappa_x, y) dy.
\end{equation}

This expression for $a'$ reflects the speed difference of turbulent structures and the travelling wave and the ensuing forcing timescale $2 \pi / \omega_{eq}$. 

The streamwise inhomogeneity of the forcing results in a three-dimensional stress state, which can be included to improve the description of the drag reduction effect. To this aim, the full spanwise shear stress $\tau_z$ is employed instead of the first derivative $\partial \tilde{w} / \partial y$ used in the YHK14 correlation. $\tau_z$ is the spanwise component of the principal stress state associated with the phase-averaged strain-rate tensor $\aver{S}_{ij}$ introduced in \S\ref{subs:QSVmod}, i.e.
\begin{equation}
\tau_z = \sum_{i=1}^3 \lambda_i \vect{e_i} \cdot \vect{\hat{z}}
\end{equation}
and was shown to correspond to $\partial \tilde{w} / \partial y$ in the OW case. 

The original prediction formula \eqref{eq:DRpred} by YHK14 can thus be modified as follows:
\begin{equation}
\Delta U = S' \left( a \tau_z|_{y^+=10}^{rms} + b \tau_z|_{y^+=15}^{rms} \right) 
- c \alpha_{y^+=12}^{rms}.
\label{eq:DRcorr}
\end{equation}
where $a,b,c$ are empirical coefficients.

The first part of the formula takes care of the drag-reduction effect and is essentially a patch to the YHK14 relation. The prefactor $S'$ introduces the dependency on $\omega_{eq}$, and vanishes in the drag increase region. The last term takes care of the drag-increase effect, by incorporating a term proportional to the root-mean-square value of the angle $\alpha_{y^+=12}$.

\begin{figure}
\centering
\includegraphics[width=0.8\textwidth]{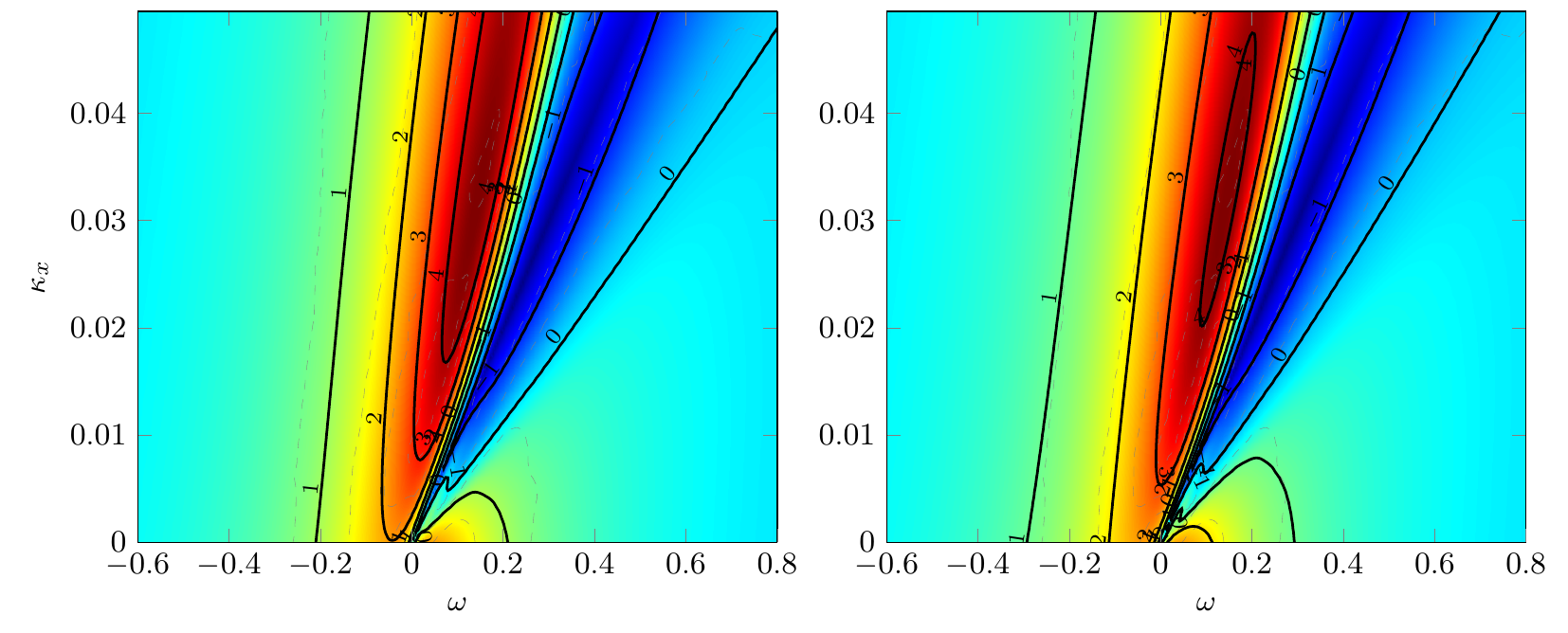}
\caption{Predicted map of drag changes: black solid contours plot predictions, while colormap and grey dashed contours are reference data from \cite{gatti-quadrio-2016}. Left: prediction by Eq.(\ref{eq:DRcorr}). Right: prediction by Eq.(\ref{eq:DRcorr2}).}
\label{fig:DU}
\end{figure}

A least square fit to the dataset available from Ref.\cite{gatti-quadrio-2016} is used to find $a=5.01$, $b=9.49$ and $c=67.46$. The resulting drag reduction map is represented in Fig.\ref{fig:DU} (left). Despite being obtained through simple observations, the prediction produces a strikingly similar map compared to the numerical results, and performs significantly better than the alternative predictive formulas illustrated in Fig.\ref{fig:corrs2}. This suggests that the most relevant aspects related to drag reduction have been taken into account. Expression \ref{eq:DRcorr} has an empirical base though, and as such can be further modified, for example by removing the assumption of linearity between drag reduction and $\tau_z$. Fig.\ref{fig:DU} (right) shows the improvement that can be obtained by assuming a non-linear variation of the drag-reducing part. After introducing an exponent $d$ to generalize Eq.(\ref{eq:DRcorr}) as:
\begin{equation}
\Delta U = a \left( S' \tau_z|_{y^+=10}^{rms} \right)^d + b \left( S' \tau_z|_{y^+=15}^{rms} \right)^d
- c \alpha_{y^+=12}^{rms}.
\label{eq:DRcorr2}
\end{equation}
the best value of $d$ that provides the best fit of the data is found at $d=0.8$, a value for which the remaining parameters become $a=4.01$, $b=8.34$ and $c=78.19$.

The minus sign in front of the last term of Eq.{\ref{eq:DRcorr} (together with the positive value of the coefficient $c$) is worth noticing, since the term represents the drag increase resulting from the effects described in \S\ref{subs:QSVmod}. Moreover, the coefficient $b$ that multiplies the shear at $y^+=15$ is positive, differently from YHK14, where it was negative. The two expressions are not immediately comparable, owing to the presence of the additional factor $S'$ in the present one. A possible interpretation is that in the OW fit by YHK14 the coefficient $b$ was modeling the decay of drag reduction occurring at low values of $\omega$, related to the enhancement of Q4 events. In the present formula, instead, the drag-increasing effect is accounted for by the additional third term, while the decay of drag reduction at low $\omega$ is enhanced by the $S^\prime$ contribution.  Moreover,  the variability of the wall-normal position of the QSV brought about by TW leads the fit to provide a weight to vortices further away from the wall: as noted commenting Fig. \ref{fig:hist}, QSV move slightly away from the wall in presence of a drag-reducing travelling wave (case TW1), hence the term $S'\tau_z|_{y^+=15}^{rms}$ represents the shear effect on structures located at larger wall distances.

\section{Conclusions}
\label{sec:conclusions}

By means of five direct numerical simulations of a turbulent channel flow at $Re_\tau=200$, this work has studied how streamwise-travelling waves for drag reduction affect quasi-streamwise vortices. The analysis of Yakeno, Hasegawa and Kasagi or YHK14 \cite{yakeno-hasegawa-kasagi-2014}, carried out for the spanwise-oscillating wall, has been extended to account for the spatially non-uniform travelling waves of spanwise velocity. Since this forcing includes the oscillating wall as a special case, this generalization provides a better view of the driving mechanism of drag reduction, and allows a clearer understanding of its main physical ingredients.

With a quadrant analysis, changes of Reynolds stresses have been related to changes of drag, to illustrate that modifications to Q2 events dominate not only drag reduction, as in the OW case, but drag increase as well. Quasi-streamwise vortices have been extracted from instantaneous flow fields via a conditional-averaging procedure based on the swirling-strength criterion; their distribution along the wall-normal direction is found to depend on the type of wall forcing. Moreover, a variation of the $y$ position of the peak for the QSV distribution con be observed for the StTW cases. 

Conditional averaging of the extracted structures has led to identify how Q2 and Q4 events change along the phase of the forcing cycle: changes are most evident with TW. Furthermore, the streamwise-dependent control generates a phase-dependent vertical bouncing of the core of Q4 and Q2 events, related to an increment of the vertical velocity component and consequently of the Reynolds stresses. The phase variation of this shift, quantified via the angle $\beta$, resembles that of the angle formed between the direction of the second eigenvector of the strain rate tensor and the spanwise direction at $y^+=12$. Indeed, the wall distance $y^+=12$ was already isolated by Duque-Daza et. al. \cite{duque-etal-2012} when studying how the growth of near-wall turbulent streaks are affected by spanwise forcing, and interpreted as the likely location of dynamically significant streaks. Moreover, it is close to $y^+=10$, the wall distance at which Touber and Leschziner \cite{touber-leschziner-2012} showed the spanwise tilting angle of low speed streaks to coincide with the angle of mean shear. The direction of the third eigenvalue of the rate-of-strain tensor has been shown to be important to capture the drag increase process.

The description of QSV and of their change along the forcing phase has been also leveraged to arrive at an improved predictive formula for drag reduction in the general case of travelling waves. The formula is based on the most recent proposal by YHK14, and extends it by amending its main points of failure. Despite its empirical nature, the formula is able to predict drag changes with surprising accuracy, thus indicating that the quantities involved in its expression play a role in the physics of drag reduction. It is confirmed that (this type of) skin friction reduction is mostly a linear phenomenon, albeit inclusion of nonlinear effects quantitatively improves the accuracy of the prediction. The presence of three empirical coefficients in the prediction formula reflects our current inability to relate the properties of the GSL with the vertical dynamics/bouncing of the QSV, which could be a direction for further work.

\begin{acknowledgements}
None.
\end{acknowledgements}

\section*{Conflict of interest}
The authors declare that they have no conflict of interest.

\end{document}